\documentclass[journal,twoside,web]{{ieeecolor}}
\usepackage{generic}
\usepackage{cite}
\usepackage{graphicx}
\usepackage{algorithm,algorithmic}
\usepackage{amsmath}
\usepackage{amssymb,amsfonts}
\usepackage{hyperref}
\usepackage{textcomp}
\usepackage{multirow}
\def\BibTeX{{\rm B\kern-.05em{\sc i\kern-.025em b}\kern-.08em
    T\kern-.1667em\lower.7ex\hbox{E}\kern-.125emX}}
\begin{document}
\title{Joint Diffusion: Mutual Consistency-Driven Diffusion Model for PET-MRI Co-Reconstruction}
\author{Taofeng Xie, Zhuo-Xu Cui, Chen Luo, Huayu Wang, Congcong Liu, Yuanzhi Zhang, Xuemei Wang,\\Yanjie Zhu, ~\IEEEmembership{Member,~IEEE}, Guoqing Chen, Dong Liang, ~\IEEEmembership{Senior Member,~IEEE}, Qiyu Jin, Yihang Zhou,\\and Haifeng Wang, ~\IEEEmembership{Senior Member,~IEEE}
\thanks{This work was partially supported by the National Natural Science Foundation of China (62271474 and 62125111), the National Key R\&D Program of China (2023YFB3811400, 2023YFC2411100 and 2021YFF0501502), the High-level Talent Program in Pearl River Talent Plan of Guangdong Province (2019QN01Y986), the Key Laboratory for Magnetic Resonance and Multimodality Imaging of Guangdong Province (2020B1212060051), the Shenzhen Science and Technology Program (KQTD20180413181834876, JCYJ20210324115810030 and KCXF20211020163408012), the “111 Project” of Higher Education Talent Training in Inner Mongolia Autonomous Region, the Inner Mongolia Medical University "ZhiYuan" Talent Program (ZY0301025), the Inner Mongolia University Independent Research Project (2022-ZZ004), the Young Talents of Science and Technology in Universities of Inner Mongolia Autonomous Region (NJYT22090), Innovative Research Team in Universities of Inner Mongolia Autonomous Region (NMGIRT2207), and the National Key Technology Research and Development Program of China (No: 2021YFF0501502).(Corresponding authors: Dong Liang; Qiyu Jin, Yihang Zhou, Haifeng Wang.)}
\thanks{Taofeng Xie and Zhuo-xu Cui contributed equally to this work.}
\thanks{Taofeng Xie, Chen Luo, Huayu Wang, Qiyu Jin, and Guoqing Chen are with School of Mathematical Sciences, Inner Mongolia University, Hohhot, China.(e-mail: tf.xie@mail.imu.edu.cn; luochen\textunderscore2019@163.com; wanghywork@gmail.com; qyjin2015@aliyun.com; cgq@imu.edu.cn)}
\thanks{Taofeng Xie is with College of Computer and Information, Inner Mongolia Medical University, Hohhot, China.}
\thanks{Zhuo-xu Cui, Yihang Zhou, and Dong Liang are with Research Center for Medical AI, Shenzhen Institutes of Advanced Technology, Chinese Academy of Sciences, Shenzhen, China.(e-mail:zx.cui@siat.ac.cn; yh.zhou@siat.ac.cn; dong.liang@siat.ac.cn)}
\thanks{Congcong Liu, Yanjie Zhu, and Haifeng Wang are with Paul C. Lauterbur Research Center for Biomedical Imaging, Shenzhen Institutes of Advanced Technology, Chinese Academy of Science, Shenzhen, China. (e-mail:cc.liu@siat.ac.cn; yj.zhu@siat.ac.cn; hf.wang1@siat.ac.cn)}
\thanks{Yuanzhi Zhang, Xuemei Wang are with Inner Mongolia Medical University Affiliated Hospital, Hohhot, China. (e-mail:dryzzhang@163.com; wangxuemei201010@163.com)}}
\maketitle
\begin{abstract}
Positron Emission Tomography and Magnetic Resonance Imaging (PET-MRI) systems can obtain functional and anatomical scans. PET suffers from a low signal-to-noise ratio. Meanwhile, the k-space data acquisition process in MRI is time-consuming. The study aims to accelerate MRI and enhance PET image quality. Conventional approaches involve the separate reconstruction of each modality within PET-MRI systems. However, there exists complementary information among multi-modal images. The complementary information can contribute to image reconstruction. In this study, we propose a novel PET-MRI joint reconstruction model employing a mutual consistency-driven diffusion mode, namely MC-Diffusion. MC-Diffusion learns the joint probability distribution of PET and MRI for utilizing complementary information. We conducted a series of contrast experiments about LPLS, Joint ISAT-net and MC-Diffusion by the ADNI dataset. The results underscore the qualitative and quantitative improvements achieved by MC-Diffusion, surpassing the state-of-the-art method.
\end{abstract}

\begin{IEEEkeywords}
joint reconstruction, positron emission tomography, magnetic resonance imaging, diffusion model, stochastic differential equations
% Enter key words or phrases in alphabetical order, separated by commas. Using the IEEE Thesaurus can help you find the best standardized keywords to fit your article. Use the thesaurus access request form for free access to the IEEE Thesaurus: \underline{https://www.ieee.org/publications/services/thesaurus-acce}\\
% \underline{ss-page.com.}
\end{IEEEkeywords}

\section{Introduction}
\label{sec:introduction}
\IEEEPARstart{M}{edical} imaging plays a pivotal role in clinical diagnosis and analysis by facilitating the visualization of anatomical and functional information. Various imaging modalities, such as Magnetic Resonance Imaging (MRI) and Computed Tomography (CT), excel in anatomical imaging, while Positron Emission Tomography (PET) and Single Photon Emission Computed Tomography (SPECT) capture functional information. The integration of both structural and functional imaging has provided invaluable scientific and diagnostic insights \cite{cherry2006multimodality}\cite{catana2013pet}. The synergy between PET and MRI is crucial for clinical diagnosis and treatment monitoring \cite{townsend2008multimodality}\cite{ehrhardt2014joint}. PET-MRI allows for the simultaneous acquisition of both functional and anatomical data \cite{catana2012pet}\cite{chen2018simultaneous}. PET and MRI scans are of the same individual, and their intrinsic characteristics are consistent. We consider PET and MRI as mutually consistent.

Structural, contrast, and resolution are disparities in PET and MRI due to their fundamentally different imaging mechanisms. Nevertheless, there are statistical correlations between PET and MRI, particularly when PET employs radiotracers like fluorodeoxyglucose (FDG).  PET images often feature lower signal-to-noise ratios (SNR) and restricted spatial resolution in visual imaging due to the inherently stochastic nature of photon emission processes \cite{oen2019image,yan2016method,lindemann2018towards,kazantsev2014novel}. In contrast, MRI offers high spatial resolution in anatomical imaging. However, the k-space data acquisition process is often deemed time-consuming. The common method reduces data acquisition for speeding. Undersampling k-space data leads to aliasing in the reconstructed images \cite{lustig2007sparse}. The joint reconstruction capitalizes on the complementary information of MRI and PET to accelerate MRI and enhance PET image quality.

Traditional joint reconstruction methods have primarily focused on structural similarity. Haber and Oldenbourg targeted geophysical applications for joint reconstruction but briefly mentioned the concept's potential relevance combined with PET and MRI reconstruction \cite{haber1997joint}. Ehrhardt et al. contributed joint reconstruction in multi-modality medical imaging first  \cite{ehrhardt2014joint}. They enhanced PET and accelerated MRI reconstructions by leveraging complementary information pertaining to structural similarity. However, multi-modal images have abundant complementary information, extending beyond mere structural similarity.

For PET and MR imaging, deep learned (DL) model-based image reconstruction techniques have emerged as highly promising methodologies, consistently surpassing the performance of conventional methods. This innovative approach enables the acquisition of prior information in the form of gradients, typically embedded within traditional reconstruction algorithms \cite{corda2020syn}. This technique involves the unrolling and interconnection of two conventional reconstruction algorithms, each corresponding to a different modality, while also learning the parameters and strengths of the regularizers. The MAPEM algorithm \cite{de1995modified} for PET and the Landweber algorithm \cite{landweber1951iteration} for MRI have been unrolled and interconnected through the regularization step. A unified prior is depicted as a U-Net architecture with shared parameters across all iterations \cite{ronneberger2015u}. The network was trained using a single-modality loss to oversee the joint reconstruction process. The method offers notable advantages, including a limited number of trainable parameters and a simpler architecture, in contrast to earlier deep-learned techniques. 

In mathematics, the joint reconstruction is known as joint inversion \cite{haber2013model}. Joint reconstruction of PET and MRI can be described as joint inversion by leveraging the information of their joint probability distribution. This distribution that serves as prior knowledge during reconstruction can enhance reconstruction quality. Consequently, joint reconstruction can be described as a data reconstruction procedure guided by this joint probability distribution. Recently, the score-based diffusion model has emerged as a state-of-the-art generative model for producing high-quality samples \cite{song2019generative}. More specifically, the score-based diffusion model accurately generates a sample by learning the probability distribution of data. This study aims to develop a joint reconstruction model that effectively harnesses the complementary information residing within both modalities, thereby significantly enhancing the quality of their respective reconstructions.
\subsection{Contributions}
Inspired by the score-based diffusion model mentioned, we propose a novel model for PET and MRI reconstruction. Our main contributions are summarized as follows:
\begin{itemize}
    \item We propose a joint reconstruction model based on the mutual consistency-driven diffusion model for PET-MRI Co-Reconstruction, termed MC-Diffusion, for the reconstruction of PET and MRI that leverage shared information effectively. The MC-Diffusion utilizes the joint probability distribution of PET and MRI in the reconstruction processes.
    \item Experiments show that the MC-Diffusion consistently outperforms conventional imaging methods and other supervised deep learning methods regarding joint reconstruction accuracy and stability.
    \item To the best of our knowledge, this work contributed to joint reconstruction by diffusion model cross-modal in medical imaging for the first time.
\end{itemize}

The paper is structured as follows: Section 2 presents related works. Section 3 details the proposed methodology. Section 4 offers the experimental results. Section 5 presents the conclusion.

\section{Related Works}
In this section, the related conventional methods, deep learning methods, and diffusion models were discussed.
\subsection{Conventional Model} 
Let $\mathbf{u}$ and $\mathbf{v}$ be PET and MRI to be reconstructed. The joint reconstruction of PET and MRI is by minimization model
$$\min\limits_{\mathbf{u,v}}F(\mathbf{u})+G(\mathbf{v})+R(\mathbf{u,v})$$
where $F(\mathbf{u})$ is the data fidelity term of PET. $G(\mathbf{v})$ is the data fidelity term of MRI. $R(\mathbf{u,v})$ is a regularization term about prior information of PET and MRI. If the regularization term $R(\mathbf{u,v})$ is separable, that is $R(\mathbf{u,v})=R(\mathbf{u})+R(\mathbf{v})$, the reconstruction of two images can be described as two independent problems.
Reconstruction of PET
$$\min\limits_{\mathbf{u}}F(\mathbf{u})+R(\mathbf{u})$$
and reconstruction of MRI
$$\min\limits_{\mathbf{v}}G(\mathbf{v})+R(\mathbf{v}).$$
The quality of reconstructed images relies on the traditional reconstruction algorithms specific to each modality. However, this approach does not fully utilize the complementary information of PET and MRI. Joint reconstruction aims to uncover complementary information of both modalities within $R(\mathbf{u,v})$ instead of segregating prior terms. Ehrhardt et al. proposed the joint reconstruction model gradient-based dependencies, particularly at finer scales operating on neighbouring voxels for enhancing the quality of PET and accelerating MRI reconstruction\cite{ehrhardt2014joint}. These studies further refined gradient-based prior models to augment joint reconstruction in live PET-MRI \cite{knoll2016joint}\cite{mehranian2017synergistic}. Parallel level sets \cite{ehrhardt2014joint} integrated gradient information from PET and MRI reconstructed images to assess the similarity between their gradients. While gradient-based priors are effective for piecewise smooth images, they may not adequately capture texture patterns within the images \cite{milanfar2012tour}\cite{cho2010content}. Zhang et al. proposed a total variation-based joint regularization with an adaptively estimated common edge indicator function as a weight \cite{zhang2018pet}. This common edge function accounts for shared structures in PET and MRI. Choi et al. proposed a tight frame-based PET-MRI joint reconstruction model leveraging the joint sparsity of tight frame coefficients \cite{choi2018pet}. However, these methods assume structural similarities between PET and MRI. The relationship between PET and MRI is complex, and conventional models can't accurately design the joint correlation between the two modalities.
\subsection{Deep Learning}
In recent years, deep learning has revolutionized medical image reconstruction, surpassing the capabilities of conventional techniques \cite{leynes2018synthetic}\cite{zhang2018pet}. Researchers employ fully convolutional neural networks (CNN) with supervised and adversarial objectives to enhance noisy, low-dose PET-MRI images \cite{dong2020deep}. The application of fully convolutional neural network image-to-image architectures in PET-MRI reconstruction has been a prevalent approach \cite{yang2020ct}. The unsupervised deep image prior has been applied to enhance PET-MRI images  \cite{gong2018pet}\cite{gong2021direct}.
Moreover, the utilization of deep learning for domain translation, primarily from MRI to CT, has yielded favourable outcomes \cite{leynes2018synthetic}. This approach has contributed to the improvement of PET-MRI, especially when integrated with traditional image reconstruction techniques.  Notable advancements include Syn-Net, a groundbreaking approach that unrolls two conventional reconstruction algorithms, integrates regularizers to learn the strengths of their respective priors and enhances the quality of reconstructed images \cite{corda2020syn}. The emergence of hybrid physics-driven DL techniques has further strengthened the PET-MRI field \cite{leynes2021attenuation}. However, it's worth noting that these DL-based methods often employ a black box design to represent the joint correlation between PET and MRI or directly map the images to ground truth.
\subsection{Diffusion Model}
% diffusion models
Diffusion models have demonstrated superior performance in various applications. Image generation models based on diffusion models include Score Matching with Langevin Dynamics (SMLD) \cite{song2019generative}, Denoising Diffusion Probabilistic Models (DDPM) \cite{ho2020denoising} and Stochastic Differential Equations (SDE)\cite{song2020score}. The framework of SDE unifies various types of diffusion models, such as SMLD and DDPM and so on, namely VE SDE, VP SDE, and sub-VP SDE. The diffusion model comprises two processes: the forward diffusion process and the reverse diffusion process. In the forward diffusion process, noise is gradually added to convert a complex data distribution into a known prior distribution. Conversely, in the reverse diffusion process, noise is removed to transform the prior distribution back into the original data distribution. The forward diffusion process is described as the solution of the following SDE
$$\mathrm{d}\mathbf{x} = f(\mathbf{x},t)dt+g(t)\mathrm{d}\mathbf{w},$$
$f(\mathbf{x},t)$ is the drift coefficient of $\mathbf{x}(t)$. $g(t)$ is the diffusion coefficient of $\mathbf{x}(t)$. $\mathbf{w}$ is the standard Wiener process. The reverse diffusion process is obtained from the following reverse-time SDE
$$\mathrm{d}\mathbf{x} = [f(\mathbf{x},t)-g(t)^{2}\boldmath{\nabla}_{\mathbf{x}}\log p_{t}(\mathbf{x})]dt+g(t)\mathrm{d}\mathbf{w}.$$
$\boldmath{\nabla}_{\mathbf{x}}\log p_{t}(\mathbf{x})$ is known as score function. $t$ is uniformly sampled over $[0,T]$. $\boldmath{\nabla}_{\mathbf{x}}\log p_{t}(\mathbf{x})$ was estimated by the score network $\mathbf{s}_{\theta}(\mathbf{x}(t),t)$. The loss function of the score network can be formulated 
\begin{equation}
    \begin{aligned}
        \theta^{*}=\underset{\theta}{\arg \min } \mathbb{E}_{t}&\left\{\lambda(t) \mathbb{E}_{\mathbf{x}(0)} \mathbb{E}_{\mathbf{x}(t) \mid \mathbf{x}(0)}\left[\left\|\mathbf{s}_{\theta}(\mathbf{x}(t), t)
        \right.\right.\right.\\&\left.\left.\left. -\boldmath{\nabla}_{\mathbf{x}(t)} \log p_{0 t}(\mathbf{x}(t) \mid \mathbf{x}(0))\right\|_{2}^{2}\right]\right\}
    \end{aligned}
    \label{SDE-loss}
\end{equation}
$\lambda(t)$ is a positive weighting function, $\theta^{*}$ denotes the optimal parameter of $\mathbf{s}_{\theta}$.

In VE-SDE framework, the noise scales $\lbrace{\sigma_{i}\rbrace}_{i=1}^{N}$ satisfies that $\sigma_{1}$ is large sufficiently, i.e., $\sigma_{max}$. $\sigma_{N}$ is small enough, i.e., $\sigma_{min}$. 
% the forward diffusion process can be expressed:
% \begin{equation*}
%     \mathrm{d}\mathbf{x} = \sqrt{\frac{d[\sigma^{2}(t)]}{dt}}\mathrm{d}\mathbf{w}
% \end{equation*}
The perturbation kernel of the VE-SDE in training  can be derived
\begin{equation}
    p_{0t}(\mathbf{x}(t)\mid \mathbf{x}(0)) = \mathcal{N}(\mathbf{x}(t);\mathbf{x}(0),[\sigma^{2}(t)-\sigma^{2}(0)]I)
    \label{perturbation kernel}
\end{equation}
By substituting Eq. (\ref{perturbation kernel}) into Eq. (\ref{SDE-loss}), the network can be trained.
The image is generated by the reverse iteration rule of 
\begin{equation}
\begin{aligned}
    \mathbf{x}_{i-1} &= \mathbf{x}_{i} + (\sigma_{i}^{2}-\sigma_{i-1}^{2})\mathbf{s}_{\theta^{*}}(\mathbf{x}_{i},i)+\sqrt{\frac{\sigma_{i-1}^{2}(\sigma_{i}^{2}-\sigma_{i-1}^{2})}{\sigma_{i}^{2}}}z_{i},\\
    &i=1,2,...,N,
    \label{continue}
\end{aligned}
\end{equation}
where $\mathbf{x}_N \sim \mathcal{N}(0,\sigma_{N}^{2}I)$, and $z_{i} \sim \mathcal{N}(0,I)$.

% \subsection{Score-based SDEs for guiding reconstruction}
% Score-based SDEs have been applied for condition generation \cite{lyu2022conversion,xie2022measurement}.
% Given the co-registered data pairs $(x_{i},y_{i})_{i=1}^{M}$, where $M$ is the number of data pairs. During the forward diffusion process, VE-SDE adds Gaussian noise with zero mean and gradually increasing variance to the data $\mathbf{x}$ and $\mathbf{y}$ is fixed. $\mathbf{x}$ can be generation through the reverse-time SDE:
% \begin{equation*}
%     \mathrm{d}\mathbf{x} = \left\{f(\mathbf{x},t)-g(t)^{2}\boldmath{\nabla}_{\mathbf{x}} \log p_{t}(\mathbf{x}, \mathbf{y})\right\}dt+g(t)\mathrm{d}\mathbf{w}.
% \end {equation*}
% Here, $\mathbf{x}$ is the image to be generated image, and $\mathbf{y}$ is the condition image.
% In the Bayesian perspective, 
% \begin{equation*}
% \begin{aligned}
% \nabla_{\mathbf{x}}\log p_{t}(\mathbf{x}\mid \mathbf{y}) 
% &=\nabla_{\mathbf{x}}(\log p_{t}\left(\mathbf{x},\mathbf{y}\right) - \log p_{t}\left(\mathbf{y})\right) \\
% &=\nabla_{\mathbf{x}}\log p_{t}\left(\mathbf{x},\mathbf{y}\right).
% \end{aligned}
% \end{equation*}
% The reverse-time SDE can be rewritten as 
% \begin{equation*}
% \mathrm{d}\mathbf{x}=\left[f(\mathbf{x},t)-g(t)g(t)^{T}\boldmath{\nabla}_{\mathbf{x}}\log p_{t}\left(\mathbf{x}, \mathbf{y}\right)\right]\mathrm{d}t+g(t)\mathrm{d}\mathbf{w}.
% \end{equation*}
% $\boldmath{\nabla}_{\mathbf{x}}\log p_{t}\left(\mathbf{x}, \mathbf{y}\right)$ can be learning in diffusion process. We can obtain the synthesis image from the condition image.

\section{Method: Mutual consistency-driven diffusion model for Joint reconstruction}
\label{Method}
\begin{figure*}[t]
    \centering
    \includegraphics[width=1\textwidth]{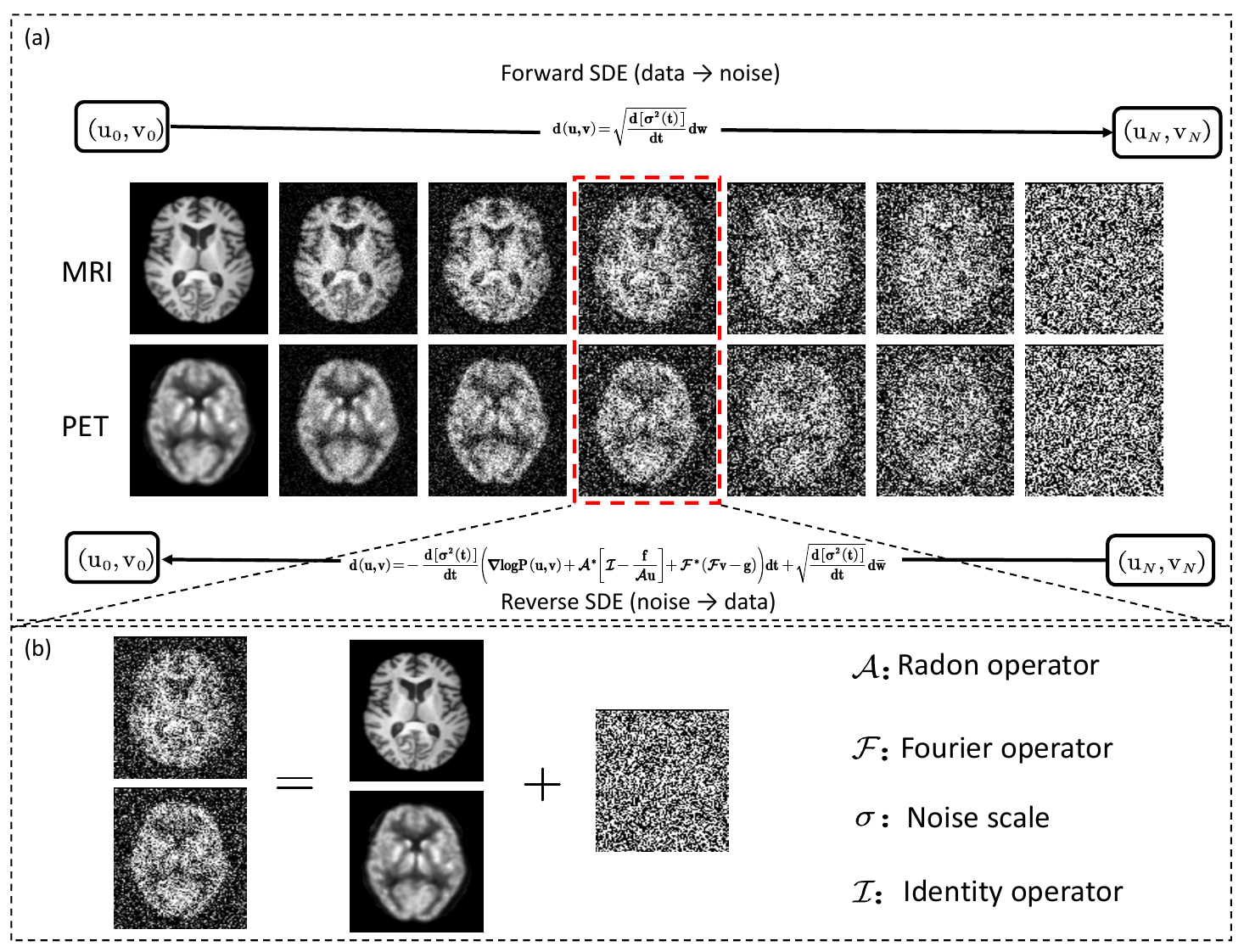}
    \caption{The scheme of MC-Diffusion. (a) In the forward process, noise at different scales is gradually added into the multi-modality and conducted Langevin MCMC sampling in the reverse process. (b) The same noise added into mutual consistency PET and MRI in the forward process. }
    \label{scheme}
\end{figure*}

This section is dedicated to introducing the problem of joint reconstruction of PET-MRI more formally. Figure \ref{scheme} shows the framework of the MC-Diffusion.
% We formulate our thoughts using probabilistic models \cite{barber2012bayesian}.
% figure*设置为图片横跨两栏
The problem of joint reconstruction can be described as 
$$\mathbf{Y} = \mathcal{D}\left(\mathbf{X}\right) + \epsilon$$
where different modalities can be written in stacked form
$$\mathbf{Y} = \begin{pmatrix}
\mathbf{f}\\
\mathbf{g}\\
\end{pmatrix},
\mathcal{D} = \begin{pmatrix}
\mathcal{A}\\
\mathcal{F}\\
\end{pmatrix},
\mathbf{X} = \begin{pmatrix}
\mathbf{u}\\
\mathbf{v}\\
\end{pmatrix},
\epsilon = \begin{pmatrix}
\varsigma\\
\xi\\
\end{pmatrix}$$
Let $\mathbf{u}$ and $\mathbf{v}$ represent the PET and MRI images to be reconstructed, respectively. $\mathbf{f}$ corresponds to the observed PET data, and $\mathbf{g}$ refers to the observed MRI data. $\mathcal{A}$ signifies the PET forward operator, which is modeled as the discrete Radon transform \cite{burger2014total}. $\mathcal{F}$ stands for the MRI forward operator, which is modeled as the discrete Fourier transform. $\epsilon$ represents the measured noise. $\varsigma$ is associated with Gaussian noise, following a normal distribution, i.e., $\varsigma \sim P(\lambda)$, and $\xi$ denotes Poisson noise, i.e., $\xi \sim \mathcal{N}(0, I)$. 

It utilizes the maximum likelihood estimated the maximum conditional probability $P(\mathbf{f,g} \mid \mathbf{u,v})$, namely
\begin{equation}
    (\mathbf{u,v})= \underset{\mathbf{u,v}}{\arg \max }P(\mathbf{u,v} \mid \mathbf{f,g})
    \label{maximum likelihood}
\end{equation}
Eq. (\ref{maximum likelihood}) is equivalent to minimizing the negative log-likelihood
\begin{equation}
    (\mathbf{u,v})=\underset{\mathbf{u,v}}{\arg \min }\left\{-\log P(\mathbf{u,v} \mid \mathbf{f,g})\right\}
    \label{log-likelihood}
\end {equation}
By Bayesian theory,
\begin{equation}
    P(\mathbf{u,v}|\mathbf{f,g}) =\frac{P(\mathbf{f,g}\mid \mathbf{u,v})P(\mathbf{u,v})}{P(\mathbf{f,g})}
    \label{Bayesian}
\end{equation}
Although $\mathbf{u}$ and $\mathbf{v}$ depend on a common object and are not independent of each other, if we have knowledge about $\mathbf{v}$ does not provide extra information about $\mathbf{f}$ and vice versa. This means, that $\mathbf{f}$ and $\mathbf{g}$ are conditionally independent given $\mathbf{u}$ and $\mathbf{v}$. Formally, this leads to the separation of the multi-modality likelihood
\begin{equation}
    P(\mathbf{f,g} \mid \mathbf{u,v}) = P(\mathbf{f} \mid \mathbf{u,v})P(\mathbf{g} \mid \mathbf{u,v}) = P(\mathbf{f} \mid \mathbf{u})P(\mathbf{g} \mid \mathbf{v})
\end{equation}
Eq. (\ref{Bayesian}) can be simplified as 
\begin{equation*}
    P(\mathbf{u,v}|\mathbf{f,g})\propto P(\mathbf{f}|\mathbf{u})P(\mathbf{g}|\mathbf{v})P(\mathbf{u,v})
\end{equation*}
Hence, Eq. (\ref{log-likelihood}) can be formulated
\begin{equation}
    \begin{aligned}
        &(\mathbf{u,v})\\=
        &\underset{(\mathbf{u,v})}{\arg \min }\left\{-\log P(\mathbf{f} \mid \mathbf{u}) - \log P(\mathbf{g} \mid \mathbf{v}) -\log P(\mathbf{u,v})\right\}
    \end{aligned}
    \label{log_1}
\end {equation}
The noise PET is commonly modelled to be Poisson \cite{ollinger1997positron}.
The PET data fidelity can be written as 
\begin{equation}
    -\log (P(\mathbf{f}|\mathbf{u})) = \sum_{i=1}^{M}\left[(\mathcal{A}\mathbf{u}_{i})-\mathbf{f}_{i}\log\left(\mathcal{A}\mathbf{u} \right)_{i}\right].
\end{equation}
The noise in MRI is commonly modelled as additive Gaussian \cite{gudbjartsson1995rician}. 
The MRI data fidelity can be written as 
\begin{equation}
     -\log(P(\mathbf{g}|\mathbf{v}))=\frac{\lambda}{2}\Vert \mathcal{F}\mathbf{v}-\mathbf{g} \Vert_{2}^{2}.
\end{equation}
Denote 
\begin{equation}
    \begin{aligned}
        &\mathcal{J}(\mathbf{u,v})\\
        =&- \log P(\mathbf{f} \mid \mathbf{u}) - \log P(\mathbf{g} \mid \mathbf{v}) - \log P(\mathbf{u},\mathbf{v})\\
        =&\sum \limits_{i}^{M}((\mathcal{A}\mathbf{u})_{i} - \mathbf{f}_{i}\log (\mathcal{A}\mathbf{u})_{i}) + \frac{1}{2\sigma^{2}}\|\mathcal{F}\mathbf{v} - \mathbf{g}\|_{2}^{2} \\&- \log P(\mathbf{u},\mathbf{v}).
    \end{aligned}
\end{equation}
Solve Eq. (\ref{log_1}) by
\begin{equation}
    \begin{aligned}
        \frac{\partial\mathcal{J}(\mathbf{u,v})}{\partial (\mathbf{u,v})}
        =&\mathcal{A}^{*}\left[\mathcal{I}-\frac{\mathbf{f}}{\mathcal{A}\mathbf{u}}\right]
        +\mathcal{F}^{*}(\mathcal{F}\mathbf{v}-\mathbf{g}) \\&- \boldmath{\nabla} \log P(\mathbf{u,v}).
    \end{aligned}
\end{equation}
In order to estimate $\boldmath{\nabla} \log P(\mathbf{u,v})$, the diffusion model can be considered. For convenience in writing, let it be denoted as  $P(\mathbf{X})=P(\mathbf{u,v})$. We can train a time-dependent score-based model $\mathbf{s}_{\theta}(\mathbf{X},t)$ such that $\mathbf{s}_{\theta}(\mathbf{X},t) \approx \boldmath{\nabla} \log P(\mathbf{X})$. 
For training estimation of the data of PET and MRI scores, we perturb the data with a small Gaussian noise. We use $N$ noise scales in the training process. 
Since $\lbrace{\sigma_{i}\rbrace}_{i=1}^{N}$ is a geometric sequence, let 
$$\sigma_{i}=\sigma(\frac{i}{N})=\sigma_{min}\left(\frac{\sigma_{max}}{\sigma_{min}}\right)^{\frac{i-1}{N-1}}, i=1,2,...,N.$$ $\sigma_{min}$ is the minimum of the noise scales satisfying $$P_{\sigma_{min}}(\mathbf{X}) \approx P(\mathbf{X}),$$ and $\sigma_{max}$ is the maximum of the noise scales. 
We choose the noise distribution to be $P_{\sigma}(\hat{\mathbf{X}}\mid \mathbf{X})=\mathcal{N}(\hat{\mathbf{X}}\mid \mathbf{X},\sigma^{2}\mathbf{I})$, then $\nabla \log P_{\sigma}(\hat{\mathbf{X}}\mid \mathbf{X})=-(\hat{\mathbf{X}}-\mathbf{X})/ \sigma^{2}$. The perturbation kernel can be written as
\begin{equation*}
    \begin{aligned}
    &P_{0t}\left(\mathbf{X}\mid \mathbf{X}_{0}\right)\\=&\mathcal{N}\left(\mathbf{X};\mathbf{X}_{0},\sigma_{min}^{2}\left(\frac{\sigma_{max}}{\sigma_{min}}\right)^{2t}\mathbf{I}\right).
    \end{aligned}
\end{equation*}
Each perturbation kernel $P_{0t}\left(\mathbf{X}\mid \mathbf{X}_{0}\right)$ corresponds to the distribution of $\mathbf{X}$ in the Markov chain
\begin{equation}
\mathbf{X}_{i}=\mathbf{X}_{i-1}+\sqrt{\sigma_{i}^{2}-\sigma_{i-1}^{2}}z_{i-1}, i=1,2,...,N
\label{Markov chain}
\end{equation}
Eq. (\ref{Markov chain}) converges to the following SDE
\begin{equation}
\mathrm{d}(\mathbf{X})= \sqrt{\frac{\mathrm{d}[\sigma^{2}(t)]}{\mathrm{d}t}}\mathrm{d}\mathbf{w}
\label{converge}
\end{equation}
% The distribution of $\mathbf{X}^{T}_{i}$ which corresponds to perturbation kernel $P_{0t}\left(\mathbf{X}^{T}\mid \mathbf{X}_{0}^{T}\right)$ is 
% \begin{equation*}
% \mathbf{X}_{i+1}^{T}= \mathbf{X}_{i}^{T} + \sigma_{min} \left( \frac{\sigma_{max}}{\sigma_{min}} \right) ^{i} z, ~ i=1,2,...,N-1,
% \end{equation*} 
$\mathbf{s}_{\theta}(\mathbf{X}, t)$ is trained via the following loss function:
\begin{equation*}
    \begin{aligned}
    L(\theta, t)=&\underset{\theta}{\arg \min } \mathbb{E}_{t}\left\{\lambda(t) \mathbb{E}_{\mathbf{X}} \mathbb{E}_{\mathbf{X} \mid \mathbf{X}_{0}}\right.\\&\left. \left[\left\|\mathbf{s}_{\theta}(\mathbf{X}, t)-\boldmath{\nabla} \log p_{0 t}(\mathbf{X} \mid \mathbf{X}_{0})\right\|_{2}^{2}\right]\right\}\\
    =&\underset{\theta}{\arg \min } \mathbb{E}_{t}\left\{\lambda(t) \mathbb{E}_{\mathbf{X}} \mathbb{E}_{\mathbf{X} \mid \mathbf{X}_{0}}\right.\\&\left. \left[\left\|\mathbf{s}_{\theta}(\mathbf{X}, t)+\frac{(\hat{\mathbf{X}}-\mathbf{X})}{\sigma^{2}}\right\|_{2}^{2}\right]\right\}
    \end{aligned}
\end{equation*}
Based on Eq. (\ref{converge}), the reconstruction process can be deduced as 
\begin{equation}
    \begin{aligned}
        &\mathrm{d}(\mathbf{u}, \mathbf{v})= -\frac{\mathrm{d}[\sigma^{2}(t)]}{\mathrm{d}t}
        \left(\boldmath{\nabla}\log P(\mathbf{u},\mathbf{v})+\mathcal{A}^{*}\left[\mathcal{I}-\frac{\mathbf{f}}{\mathcal{A}\mathbf{u}}\right]\right.\\& \left.+\mathcal{F}^{*}(\mathcal{F}\mathbf{v}-\mathbf{g})\right)\mathrm{d}t+\sqrt{\frac{\mathrm{d}[\sigma^{2}(t)]}{\mathrm{d}t}}
        \mathrm{d}\mathbf{\bar{w}}
    \end{aligned}
\end{equation}
The reconstruction process is the predictor and corrector are executed alternately. The predictor can be described as 
% $(\mathrm{u}_{0},\mathrm{v}_{0})$
% $(\mathrm{u}_{N},\mathrm{v}_{N})$
\begin{equation*}
    \begin{aligned}
        \mathbf{X}_{i}= &\mathbf{X}_{i+1}-\mathbf{s}_{\theta}(\mathbf{X}_{i+1},t)-\mathcal{A}^{*}\left[\mathcal{I}-\frac{\mathbf{f}_{i+1}}{\mathcal{A}\mathbf{u}_{i+1}}\right]
        \\ &-\mathcal{F}^{*}(\mathcal{F}\mathbf{v}_{i+1}-\mathbf{g}_{i+1})+z
    \end{aligned}
\end{equation*}
The corrector can be described as 
\begin{equation*}
    \begin{aligned}
        \mathbf{X}_{i} = &\mathbf{X}_{i+1}+\eta_{i} \mathbf{s}_{\theta}(\mathbf{X}_{i+1},t) +\mathcal{A}^{*}\left[\mathcal{I}-\frac{\mathbf{f}_{i+1}}{\mathcal{A}\mathbf{u}_{i+1}}\right]
         \\ &+\mathcal{F}^{*}(\mathcal{F}\mathbf{v}_{i+1}-\mathbf{g}_{i+1})+\sqrt{2\eta_{i}} z
    \end{aligned}
\end{equation*}

We provide the joint sampling algorithm in Algorithm \ref{alg: PC Sampling-JR_SDE}.
\begin{algorithm}
    \caption{Joint Sampling in the reverse process.}
    \label{alg: PC Sampling-JR_SDE}
    \begin{algorithmic}[1]
	\REQUIRE $\{\sigma_i\}_{i=1}^N, \mathbf{f}, \mathbf{g},\mathbf{u}, \mathbf{v}, \epsilon_{1}, \epsilon_{2}, \mu_{1}, \mu_{2}, \alpha_{1}, \alpha_{2}, \beta_{1}, \beta_{2}, \rho_{1}, $\\$\rho_{2}, \mathcal{M}_u, N, M$. $\mathcal{M}_u$ is the undersampling mask of MRI \\
        % \Ensure $x_0^0,y_0^0$
	% \STATE{$\mathbf{u}_{N} \sim \mathcal{N}(0, I)$}\\
 %        \STATE{$\mathbf{v}_{N} \sim \mathcal{N}(0, I)$}\\
	   % \State{Initialize $\tilde{\bfx}_0$}
	    \textbf{For} {$i = N-1$ to 0}
	        \STATE \hspace{0.5cm}{$\mathbf{z} \sim \mathcal{N}(0, I)$}
	        \STATE \hspace{0.5cm}{$(\mathbf{s}_{P},\mathbf{s}_{M}) \leftarrow \mathbf{s}_{\theta^*}\left(\mathbf{u}_{i+1}, \mathbf{v}_{i+1}, i+1\right)$}
            \STATE \hspace{0.5cm}{$\mathbf{G}_{P}= \mathbf{Fbp}\left(\mathcal{A}(\mathbf{u}_i) - \mathbf{f}_{i}\right)$}
	        \STATE \hspace{0.5cm}{$\mathbf{G}_{M}= \mathcal{F}^{*}\left(\mathcal{F}(\mathbf{u}_i) - \mathbf{g}_{i+1}\right) \cdot \mathcal{M}_u$} 
	        \STATE \hspace{0.5cm}{$\epsilon_{1} \leftarrow \lambda_1\left(\|\mathbf{s}_{P}\|_{2} /\|\mathbf{G}_{P}\|_{2}\right)$}
            \linespread{1}
            \STATE \hspace{0.5cm}{$\epsilon_{2} \leftarrow \lambda_2\left(\|\mathbf{s}_{M}\|_{2} /\|\mathbf{G}_{M}\|_{2}\right)$}
            \linespread{1}
	        \STATE \hspace{0.5cm}\parbox[t]{0.8\linewidth}{$\mathbf{u}_{i} \leftarrow \mathbf{u}_{i+1}+\left(\sigma_{i+1}^{2}-\sigma_{i}^{2}\right)(\mathbf{s}_{P}-\epsilon_{1}\mathbf{G}_{P}) +\sqrt{\sigma_{i+1}^{2}-\sigma_{i}^{2}}\mathbf{z}$}
            \STATE \hspace{0.5cm}\parbox[t]{0.8\linewidth}{$\mathbf{v}_{i} \leftarrow \mathbf{v}_{i+1}+\left(\sigma_{i+1}^{2}-\sigma_{i}^{2}\right)(\mathbf{s}_{M}-\epsilon_{2}\mathbf{G}_{M})+\sqrt{\sigma_{i+1}^{2}-\sigma_{i}^{2}}\mathbf{z}$}\\
             \hspace{0.5cm}\textbf{For} {$j \gets 1$ to $M$}
                \STATE \hspace{1.0cm}{$\mathbf{z} \sim \mathcal{N}(0, I)$}
                \STATE \hspace{1.0cm}{$\left(\mathbf{s}_{P},\mathbf{s}_{M}\right) \leftarrow \mathbf{s}_{\theta^*}\left(\mathbf{u}_{i}^{j-1}, \mathbf{v}_{i}^{j-1}, i\right)$}
                \STATE \hspace{1.0cm}{$\mathbf{G}_{P}= \mathbf{Fbp}\left(\mathcal{A}(\mathbf{u}_{i}^{j-1}) - \mathbf{f}_{i}\right)$}
	          \STATE \hspace{1.0cm}{$\mathbf{G}_{M}= \mathcal{F}^{*}\left(\mathcal{F}(\mathbf{v}_i) - \mathbf{g}_{i}\right) \cdot \mathcal{M}_u$} 
                \STATE \hspace{1.0cm}{$\mu_{1} \leftarrow 2 \alpha_{1}\left(r_{1}\|\mathbf{z}\|_{2} /\|s_{P}\|_{2}\right)^{2}$}
                \STATE \hspace{1.0cm}{$\rho_{1} \leftarrow \beta_{1}\left(\|\mathbf{s}_{P}\|_{2} /\|\mathbf{G}_{P}\|_{2}\right)$}
                \STATE \hspace{1.0cm}{$\mu_{2} \leftarrow 2 \alpha_{2}\left(r_{2}\|\mathbf{z}\|_{2} /\|\mathbf{s}_{M}\|_{2}\right)^{2}$}
                \STATE \hspace{1.0cm}{$\rho_{2} \leftarrow \beta_{2}\left(\|\mathbf{s}_{M}\|_{2} /\|\mathbf{G}_{M}\|_{2}\right)$}
                \STATE \hspace{1.0cm}{$\mathbf{u}_{i}^{k} \leftarrow \mathbf{u}_{i}^{k-1}+\mu_{1} (\mathbf{s}_{P}-\rho_{1}\mathbf{G}_{P})+\sqrt{2 \mu_{1}} \mathbf{z}$}
                \STATE \hspace{1.0cm}{$\mathbf{v}_{i}^{k} \leftarrow \mathbf{v}_{i}^{k-1}+\mu_{2} (\mathbf{s}_{M}-\rho_{2}\mathbf{G}_{M})+\sqrt{2 \mu_{2}} \mathbf{z}$}\\
            \hspace{0.5cm}\textbf{End For}
            % \STATE \hspace{0.5cm}{$\mathbf{u}_{i-1}^{0} \leftarrow \mathbf{u}_{i}^{M}$}
            % \STATE \hspace{0.5cm}{$\mathbf{v}_{i-1}^{0} \leftarrow \mathbf{v}_{i}^{M}$}
            \\
        \textbf{End For}\\
        \textbf{Return} {${\mathbf{u}_{0}^{0}},{\mathbf{v}_{0}^{0}}$}
    \end{algorithmic}
\end{algorithm}
\section{Experiments}
\subsection{Experimental Details} 
% 实验数据
\subsubsection{Datasets}
% We perform the joint reconstruction on the synthesized images of PET and MRI of size $128 \times 128$. The PET imaging operator $\mathcal{A}$ is composed of X-ray transform as in \cite{zhang2020metainv} and a one dimensional Gaussian kernel along radial direction on model the lack of resolution of the operator. To generate the MRI data, we perform subsampling pattern of cartesian.
Our training dataset was sourced from the Alzheimer's Disease Neuroimaging Initiative (ADNI) dataset, comprising 167 individuals with a total of 5010 image pairs \cite{jack2008alzheimer}. The test set was conducted on a single individual, resulting in 30 image pairs.
Data preprocessing for MRI consisted of three key steps. Firstly, we performed anterior commissure (AC) and posterior commissure (PC) alignment using Statistical Parametric Mapping (SPM)\footnote{https://www.fil.ion.ucl.ac.uk/spm}. 
Secondly, the elimination of non-brain tissue was carried out by HD-BET \cite{schell2019automated} to remove any irrelevant information. Additionally, non-brain tissue was similarly removed from the PET images using the MRI-based non-brain tissue mask. Finally, the MRI and PET images of each subject were spatially aligned. Each MRI scan was brought into alignment with the standardized Montreal Neurological Institute (MNI) coordinate space, while the PET scan for the same subject was aligned to the MRI using FMRIB's Software Library (FSL)\footnote{https://fsl.fmrib.ox.ac.uk/fsl}. MRI and PET images were reshaped to a volume of $109 \times 109 \times 91$ voxels by aligning. Each MRI and PET volume consisted of 2D axial image slices, and each slice of MRI and PET was resampled to $128 \times 128$. In our experiments, we focused on slices ranging from the 30th to the 60th axial slice of each subject. To model the lack of resolution, the sinogram of size is  $128 \times 300$ in PET projection images. In all experiments, missing sinograms of size were the same.
\subsubsection{Parameter Configuration}
We conducted a comparative analysis between JP-SDE, the linear parallel level sets (LPLS) method \cite{ehrhardt2014joint}, and the supervised deep learning method (Joint ISTA-Net)  \cite{yang2020model}, with the results of each method tuned to the best. In the training stage, Joint ISTA-Net, and MC-Diffusion were trained with 500 epochs. MC-Diffusion controlled the noise level in forward diffusion by setting $\sigma_{max}=348$ and $\sigma_{min}=0.1$ in Algorithm \ref{alg: PC Sampling-JR_SDE}. In the above experiments, the noise scale N was 1000. It means 1000 iterations were needed in the reverse-time process for sample generation, taking an average of 4 minutes to reconstruct $128 \times 128$ PET and MRI on an NVIDIA V100 GPU.
\subsubsection{Performance Evaluation}
We employed three quantitative metrics to assess the reconstruction performance: normalized mean square error (NMSE), peak signal-to-noise ratio (PSNR), and the structural similarity (SSIM) index \cite{wang2004image}. A smaller NMSE, and larger values of PSNR and SSIM, are indicative of superior reconstruction quality.
\subsection{Experimental Results}
In this section, we aim to evaluate the effectiveness of our proposed method, JD-SDE, and substantiate our claims through a series of experiments. Initially, we will compare our approach with the traditional method, LPLS, to highlight the advantages of our method. Subsequently, we will make a comparative analysis with the deep learning method, the Joint ISTA method, and demonstrate the superior performance of our approach.
% This subsection shows the reconstruction results of the above methods. score-based SDEs are unsupervised learning methods in which only the fully sampled images are used for network training without the paired undersampled data. Therefore, no undersampling pattern is needed for network training in VE-SDEs. 
\subsubsection{Ablation Study}
\begin{figure}[H]
    \centering
    \includegraphics[width=0.45\textwidth]{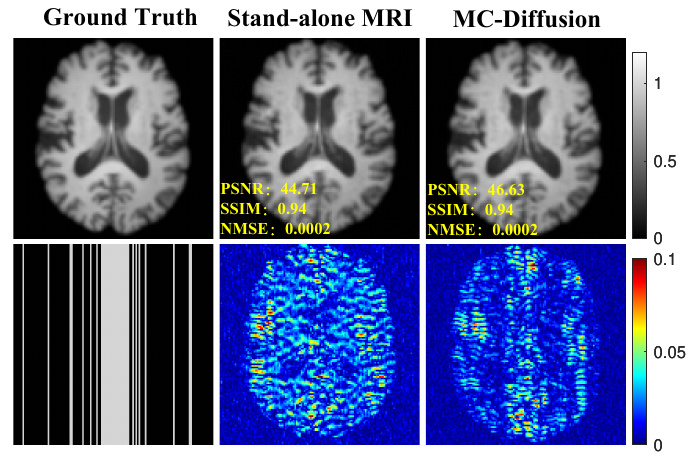}
    \caption{Reconstruction results under cartesian undersampling at 4-fold. The values in the corner are each slice's PSNR/SSIM/NMSE values. The first row describes the ground truth of MRI and the results of reconstruction by independent model and joint reconstruction model. The second row shows MRI undersampling patterns and error views. The grayscale of the reconstructed images and the error images' colour bar are on the figure's right.
   }
    \label{MRI}
\end{figure}

\begin{figure}[H]
    \centering
    \includegraphics[width=0.45\textwidth]{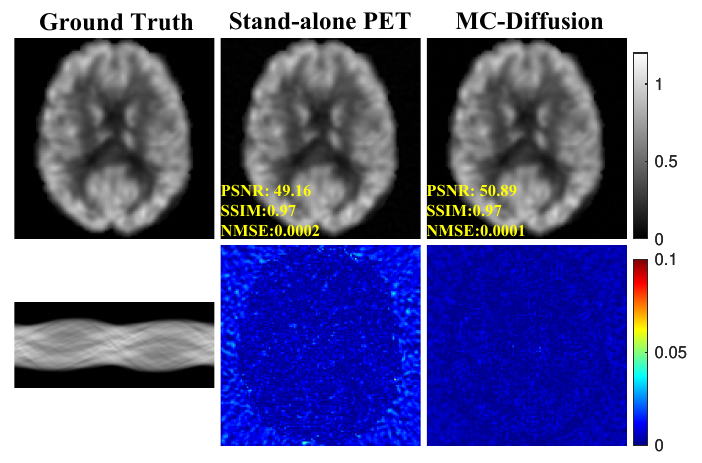}
    \caption{Reconstruction results for sinograms of size $128 \times 300$. The first row describes the ground truth of PET and the results of reconstruction by independent model and joint reconstruction model. The second row shows PET sinogram data and error views. }
    \label{PET}
\end{figure}

In this section, we will conduct ablation experiments to affirm the efficacy of this complementary information.  In Figure \ref{MRI}, "Stand-alone" refers to individual reconstruction using the diffusion model with a Cartesian undersampling factor of 4. A comparison of the results from single reconstruction MRI and joint reconstruction MRI reveals that the lack of complementary information significantly compromises the quality of the reconstruction results. Thus, the presence of complementary information intermodal is evident. The quantitative metrics presented in Table \ref{single mri table} correspondingly validate the consistency of the performance with visual perception. Figure \ref{PET} illustrates the result of PET reconstruction in an ablation experiment. The results of the PET ablation experiment mirror those of the MRI ablation experiment, indicating that complementary information contributes to PET reconstruction. The quantitative metrics of PET reconstruction are presented in Table \ref{single pet table}. This ablation experiment effectively validates the efficacy of the complementary information. It is apparent that the joint reconstruction model consistently outperforms the individual reconstruction model, confirming the existence of a correlation between the two modality images. Exploiting this correlation leads to improved reconstruction results.

\begin{table}
\centering
\caption{Quantitative comparison of MRI reconstruction for ablation studies. Highlighting the best results in bold.}
\label{single mri table}
\setlength{\tabcolsep}{7pt}
\begin{tabular}{c|c c}
\hline
\hline
& Stand-alone MRI& MC-Diffusion \\
\hline
{PSNR}& $41.5485\pm 4.5298$&$\textbf{42.5714}\pm \textbf{3.3082}$ \\
{SSIM}& $0.9456\pm 0.0359$&$\textbf{0.9554}\pm \textbf{0.0359}$ \\
{NMSE}& $0.0004\pm 0.0004$& $\textbf{0.0002}\pm \textbf{0.0003}$\\
\hline
\hline
\end{tabular}
\end{table}
\begin{table}
\centering
\caption{Quantitative comparison of PET reconstruction for ablation studies. Highlighting the best results in bold.}
\label{single pet table}
\setlength{\tabcolsep}{7pt}
\begin{tabular}{c|c c}
\hline
\hline
&Stand-alone PET& MC-Diffusion \\
\hline
{PSNR}& $43.9976\pm 2.2171$& $\textbf{50.7768}\pm \textbf{4.4200}$\\
{SSIM}& $0.9662\pm 0.0386$& $\textbf{0.9732}\pm \textbf{0.0364}$\\
{NMSE}& $0.0001\pm 0.00007$&  $\textbf{0.0001}\pm \textbf{0.0012}$\\
\hline
\hline
\end{tabular}
\end{table}
\subsubsection{Comparison Experiment}
To illustrate the effectiveness of the proposed method, a comprehensive series of comparative experiments was carried out in this section. In particular, we compared our method to the traditional approach, linear parallel level sets (LPLS) \cite{ehrhardt2014joint}, and the supervised deep learning method known as joint ISTA-Net \cite{yang2020model}.

\begin{figure}[H]
    \centering
    \includegraphics[width=0.45\textwidth]{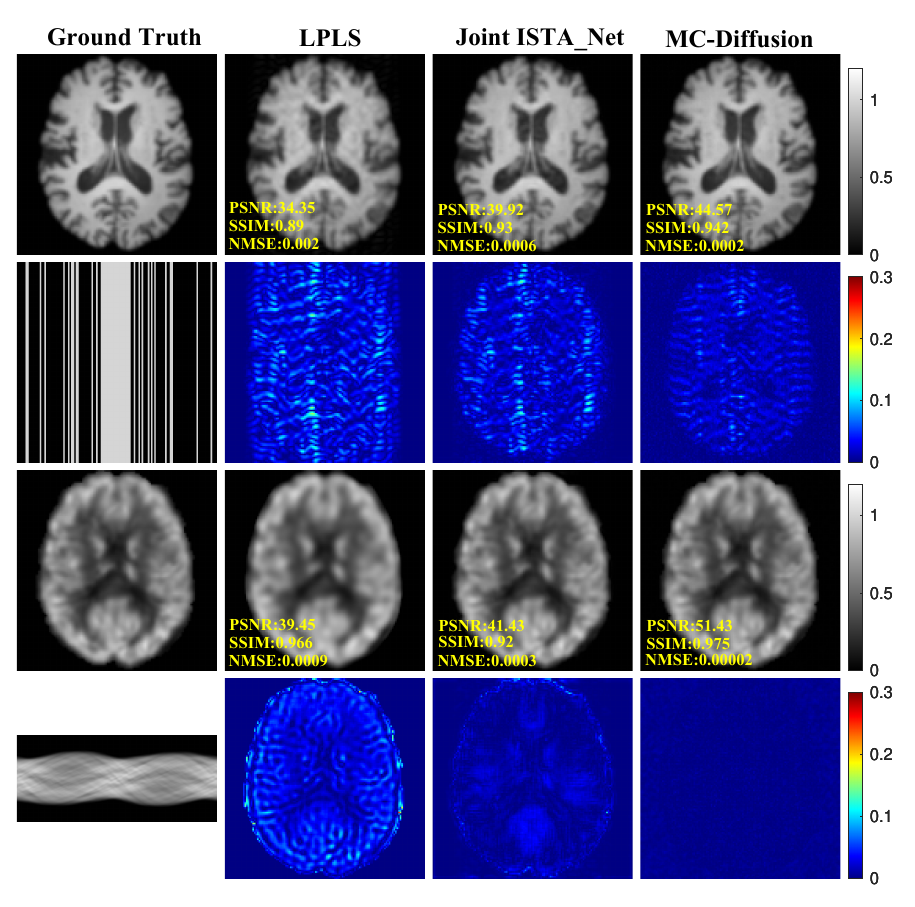}
    \caption{Joint reconstruction results under cartesian undersampling at 3-fold and for sinograms of size $128 \times 300$. The values in the corner are each slice's PSNR/SSIM/NMSE values. The first row describes the ground truth of MRI and the results of reconstruction by contrast models. The second row shows MRI undersampling patterns and error views. The third row describes the ground truth of PET and the results of joint reconstruction by contrast models. The fourth row shows PET sinogram data and error views. The grayscale of the reconstructed images and the error images' colour bar are on the figure's right.}
    \label{3acc}
\end{figure}

Figure \ref{3acc} illustrates the reconstruction outcomes of various methods with a Cartesian undersampling factor of 3. The results clearly show that MRI reconstructed using the LPLS method exhibits severe artifacts, while Joint ISTA Net introduces aliasing patterns in its reconstructions. Table \ref{3acc_table} complements these visual observations with quantitative metrics, further validating the effectiveness of our proposed approach.

\begin{figure}[H]
    % \centering
    \includegraphics[width=0.45\textwidth]{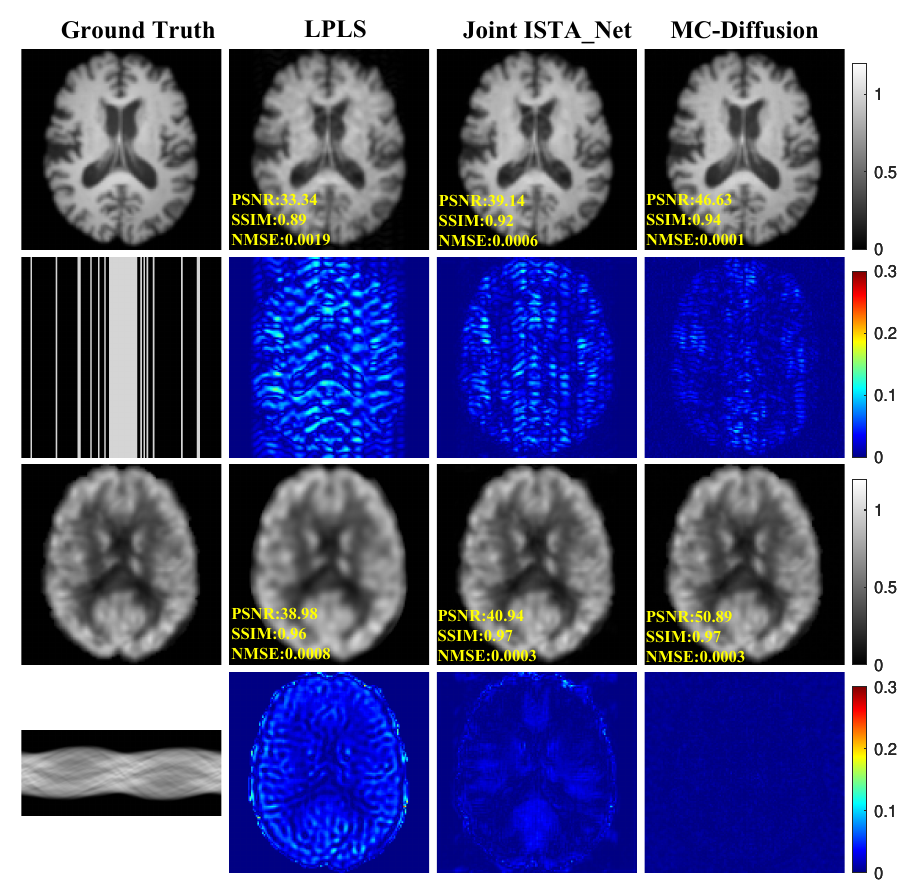}
    \caption{Joint reconstruction results under cartesian undersampling at 4-fold and for sinograms of size $128 \times 300$. The first row describes the ground truth of MRI and the results of reconstruction by contrast models. The second row shows MRI subsampling patterns and error views. The third row describes the ground truth of PET and the results of joint reconstruction by contrast models. The fourth row shows PET sinogram data and error views.}
    \label{4acc}
\end{figure}

\begin{figure}[H]
    \centering
    \includegraphics[width=0.45\textwidth]{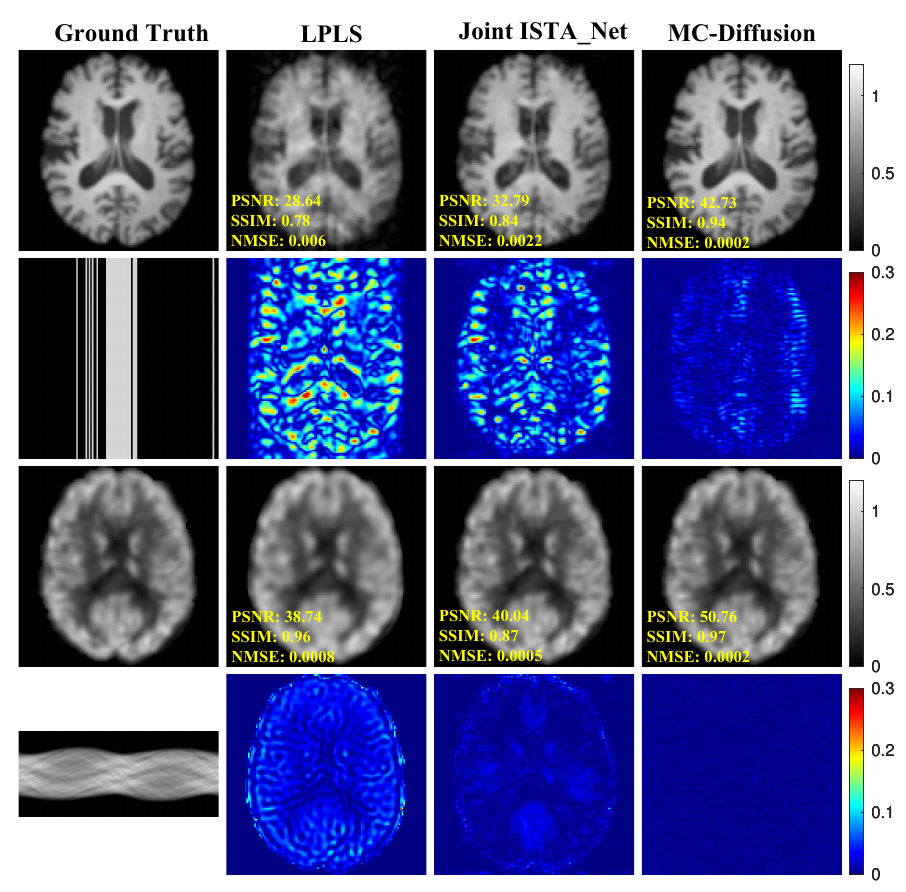}
    \caption{Joint reconstruction results under cartesian undersampling at 5-fold and for sinograms of size $128 \times 300$. The first row describes the ground truth of MRI and the results of reconstruction by contrast models. The second row shows MRI subsampling patterns and error views. The third row describes the ground truth of PET and the results of joint reconstruction by contrast models. The fourth row shows PET sinogram data and error views.}
    \label{5acc}
\end{figure}

Similar experiments were conducted with Cartesian undersampling factor of 4 and Cartesian undersampling factor of 5 for MRI, and the same undersampling model was applied to PET in all cases. Figure \ref{4acc} displays the reconstruction outcomes for a Cartesian undersampling factor of 4, and Table \ref{4acc_table} provides quantitative metrics to reinforce the effectiveness of our proposed approach. Similarly, Figure \ref{5acc} shows the results for a Cartesian factor of 5, with Table \ref{5acc_table} presenting quantitative metrics.

The stability of PET reconstruction remains consistent regardless of variations in MRI undersampling transformations. In contrast, the quality of MRI reconstruction varies with the degree of undersampling; more pronounced undersampling leads to diminished reconstruction quality. The edges in the PET image reconstructed using the joint reconstruction method are well captured. Compared to existing joint reconstruction models, our proposed models exhibit fewer artifacts in both modality images, resulting in visual improvements over the current joint reconstruction methods.
Table \ref{3acc_table} presents quantitative metrics that compare the reconstruction results of all methods using 3-fold MRI and PET undersampling. It is evident that the proposed method outperforms the comparison methods by a significant margin in terms of the quantitative metrics. This observation underscores the effectiveness of the proposed method in addressing the joint reconstruction problem.
\begin{table}
\centering
\caption{Quantitative comparison with 3-fold MRI. Highlighting the best results in bold.}
\centering
\label{3acc_table}
\setlength{\tabcolsep}{3pt}
\begin{tabular}{c|c c c}
\hline
\hline
& LPLS& Joint ISTA-Net& MC-Diffusion \\
\hline
\multirow{3}{*}{PET}& $38.4944\pm 5.8396$&$40.3399\pm 3.9571$ &$\textbf{51.3938}\pm \textbf{2.3749}$ \\
& $0.9652\pm 0.0311$& $0.9727\pm 0.0322$&$\textbf{0.9750}\pm \textbf{0.0352}$ \\
& $0.0009\pm 0.0011$& $0.0003\pm 0.0002$&$\textbf{0.00002}\pm \textbf{0.00001}$ \\
\hline
\multirow{3}{*}{MRI}& $33.3756\pm 3.4903$& $38.8761 \pm 4.5036$&$\textbf{43.9146}\pm \textbf{4.0141}$ \\
& $0.8985\pm 0.0374$&$0.9246\pm 0.0321$&$\textbf{0.9580}\pm \textbf{0.0337}$ \\
& $0.0020\pm 0.0017$&$0.0006\pm 0.0007$& $\textbf{0.0001}\pm \textbf{0.0001}$\\
\hline
\hline
\end{tabular}
\end{table}
Figure \ref{4acc} depict the joint reconstruction results with 4-fold MRI undersampling. The images were reconstructed using LPLS, Joint-ISTA-Net, and MC-Diffusion. The average quantitative metrics for the ADNI dataset are presented in Table \ref{4acc_table}. MC-Diffusion achieves the highest PSNR and SSIM in all the compared methods. Similarly, MC-Diffusion attains the lowest NMSE in all the compared methods.
\begin{table}
\caption{The average quantitative metrics at 4-fold. Highlighting the best results in bold.}
\centering
\label{4acc_table}
\setlength{\tabcolsep}{3pt}
\begin{tabular}{c|c c c}
\hline
\hline
& LPLS& Joint ISTA-Net& MC-Diffusion \\
\hline
\multirow{3}{*}{PET}& $38.4898\pm 5.8315$& $40.9373\pm 3.7957$& $\textbf{50.7768}\pm \textbf{4.4200}$\\
& $0.9637\pm 0.0322$& $0.9706\pm 0.0351$& $\textbf{0.9732}\pm \textbf{0.0364}$\\
& $0.0008\pm 0.0011$& $0.0003\pm 0.0002$& $\textbf{0.0001}\pm \textbf{0.0012}$\\
\hline
\multirow{3}{*}{MRI}& $33.3311\pm 2.9657$&$38.1805\pm 3.6424$ &$\textbf{42.5714}\pm \textbf{3.3082}$ \\
& $0.8946\pm 0.0500$& $0.9206\pm 0.0378$&$\textbf{0.9456}\pm \textbf{0.0359}$ \\
& $0.0019\pm 0.0013$&$0.0006\pm 0.00005$& $\textbf{0.0002}\pm \textbf{0.0003}$\\
\hline
\hline
\end{tabular}
\end{table}
The results of joint reconstruction with 5-fold MRI undersampling are visually represented in Figure \ref{5acc}. Reconstruction was carried out using LPLS, Joint-ISTA-Net, and MC-Diffusion. Table \ref{5acc_table} compiles the average quantitative metrics. MC-Diffusion outperforms all the compared methods by achieving the highest PSNR and SSIM. Additionally, MC-Diffusion demonstrates the lowest NMSE among the methods in comparison.
\begin{table}
\caption{Comparison of PSNR, SSIM, and NMSE at 5-fold. Highlighting the best results in bold.}
\centering
\label{5acc_table}
\setlength{\tabcolsep}{3pt}
\begin{tabular}{c|c c c}
\hline
\hline
& LPLS& Joint ISTA-Net& MC-Diffusion \\
\hline
\multirow{3}{*}{PET}& $38.8059\pm 5.7741$& $40.0105\pm 4.7626$&$\textbf{50.7618}\pm \textbf{4.3857}$\\
& $0.9652\pm 0.0311$& $0.9707\pm 0.0336$&$\textbf{0.9732}\pm \textbf{0.0365}$ \\
& $0.0008\pm 0.0011$&$0.0005\pm 0.0011$&$\textbf{0.0002}\pm \textbf{0.0013}$ \\
\hline
\multirow{3}{*}{MRI}& $27.7578\pm 2.2717$& $32.3583\pm 2.4783$& $\textbf{42.2932}\pm \textbf{.9325}$\\
& $0.7876\pm 0.0421$& $0.8453\pm 0.0328$& $\textbf{0.9434}\pm \textbf{0.0359}$\\
& $0.0062\pm 0.0031$& $0.0022\pm 0.0016$& $\textbf{0.0002}\pm \textbf{0.0002}$\\
\hline
\hline
\end{tabular}
\end{table}

\section{Conclusions}
In this paper, we propose a novel joint diffusion model for the joint reconstruction of PET and MRI. Difference from traditional reconstruction models, the joint reconstruction model presented in this paper capitalizes on the synergistic potential of complementary information across multiple modalities. This method allows for a more accurate characterization of the joint distribution between PET and MRI and utilizes it as a joint prior to generating clean images from undersampling data. The numerical experiments have demonstrated that MC-Diffusion outperforms the parallel imaging, and supervised deep learning in terms of joint reconstruction accuracy. This performance gain of our proposed models mainly comes from taking the accurate learning of joint probability distributions of PET and MRI. However, we utilized the U-Net architecture from \cite{lin2017refinenet} for estimating the score function in the diffusion model. Given the diversity of U-Net architectures, we refrained from attempting other network structures. This decision was primarily motivated by the focus of our research, which is the model's capability to learn joint probability distributions rather than conducting an exhaustive investigation of network architectures. This is our further work.
\section*{References}
% \bibliography{refs}
\bibliographystyle{IEEEtran}
\bibliography{IEEEabrv,main }

\begin{thebibliography}{10}
\providecommand{\url}[1]{#1}
\csname url@samestyle\endcsname
\providecommand{\newblock}{\relax}
\providecommand{\bibinfo}[2]{#2}
\providecommand{\BIBentrySTDinterwordspacing}{\spaceskip=0pt\relax}
\providecommand{\BIBentryALTinterwordstretchfactor}{4}
\providecommand{\BIBentryALTinterwordspacing}{\spaceskip=\fontdimen2\font plus
\BIBentryALTinterwordstretchfactor\fontdimen3\font minus \fontdimen4\font\relax}
\providecommand{\BIBforeignlanguage}[2]{{%
\expandafter\ifx\csname l@#1\endcsname\relax
\typeout{** WARNING: IEEEtran.bst: No hyphenation pattern has been}%
\typeout{** loaded for the language `#1'. Using the pattern for}%
\typeout{** the default language instead.}%
\else
\language=\csname l@#1\endcsname
\fi
#2}}
\providecommand{\BIBdecl}{\relax}
\BIBdecl

\bibitem{cherry2006multimodality}
S.~R. Cherry, ``Multimodality in vivo imaging systems: twice the power or double the trouble?'' \emph{Annu. Rev. Biomed. Eng.}, vol.~8, pp. 35--62, 2006.

\bibitem{catana2013pet}
C.~Catana, A.~R. Guimaraes, and B.~R. Rosen, ``Pet and mr imaging: the odd couple or a match made in heaven?'' \emph{Journal of Nuclear Medicine}, vol.~54, no.~5, pp. 815--824, 2013.

\bibitem{townsend2008multimodality}
D.~Townsend, ``Multimodality imaging of structure and function,'' \emph{Physics in Medicine \& Biology}, vol.~53, no.~4, p.~R1, 2008.

\bibitem{ehrhardt2014joint}
M.~J. Ehrhardt, K.~Thielemans, L.~Pizarro, D.~Atkinson, S.~Ourselin, B.~F. Hutton, and S.~R. Arridge, ``Joint reconstruction of pet-mri by exploiting structural similarity,'' \emph{Inverse Problems}, vol.~31, no.~1, p. 015001, 2014.

\bibitem{catana2012pet}
C.~Catana, A.~Drzezga, W.-D. Heiss, and B.~R. Rosen, ``Pet/mri for neurologic applications,'' \emph{Journal of nuclear medicine}, vol.~53, no.~12, pp. 1916--1925, 2012.

\bibitem{chen2018simultaneous}
Z.~Chen, S.~D. Jamadar, S.~Li, F.~Sforazzini, J.~Baran, N.~Ferris, N.~J. Shah, and G.~F. Egan, ``From simultaneous to synergistic mr-pet brain imaging: A review of hybrid mr-pet imaging methodologies,'' \emph{Human brain mapping}, vol.~39, no.~12, pp. 5126--5144, 2018.

\bibitem{oen2019image}
S.~K. {\O}en, L.~B. Aasheim, L.~Eikenes, and A.~M. Karlberg, ``Image quality and detectability in siemens biograph pet/mri and pet/ct systems—a phantom study,'' \emph{EJNMMI physics}, vol.~6, pp. 1--16, 2019.

\bibitem{yan2016method}
J.~Yan, J.~Schaefferkoetter, M.~Conti, and D.~Townsend, ``A method to assess image quality for low-dose pet: analysis of snr, cnr, bias and image noise,'' \emph{Cancer Imaging}, vol.~16, no.~1, pp. 1--12, 2016.

\bibitem{lindemann2018towards}
M.~E. Lindemann, V.~Stebner, A.~Tschischka, J.~Kirchner, L.~Umutlu, and H.~H. Quick, ``Towards fast whole-body pet/mr: Investigation of pet image quality versus reduced pet acquisition times,'' \emph{PLoS One}, vol.~13, no.~10, p. e0206573, 2018.

\bibitem{kazantsev2014novel}
D.~Kazantsev, S.~Ourselin, B.~F. Hutton, K.~J. Dobson, A.~P. Kaestner, W.~R. Lionheart, P.~J. Withers, P.~D. Lee, and S.~R. Arridge, ``A novel technique to incorporate structural prior information into multi-modal tomographic reconstruction,'' \emph{Inverse Problems}, vol.~30, no.~6, p. 065004, 2014.

\bibitem{lustig2007sparse}
M.~Lustig, D.~Donoho, and J.~M. Pauly, ``Sparse mri: The application of compressed sensing for rapid mr imaging,'' \emph{Magnetic Resonance in Medicine: An Official Journal of the International Society for Magnetic Resonance in Medicine}, vol.~58, no.~6, pp. 1182--1195, 2007.

\bibitem{haber1997joint}
E.~Haber and D.~Oldenburg, ``Joint inversion: a structural approach,'' \emph{Inverse problems}, vol.~13, no.~1, p.~63, 1997.

\bibitem{corda2020syn}
G.~Corda-D'Incan, J.~A. Schnabel, and A.~J. Reader, ``Syn-net for synergistic deep-learned pet-mr reconstruction,'' in \emph{2020 IEEE Nuclear Science Symposium and Medical Imaging Conference (NSS/MIC)}.\hskip 1em plus 0.5em minus 0.4em\relax IEEE, 2020, pp. 1--5.

\bibitem{de1995modified}
A.~R. De~Pierro, ``A modified expectation maximization algorithm for penalized likelihood estimation in emission tomography,'' \emph{IEEE transactions on medical imaging}, vol.~14, no.~1, pp. 132--137, 1995.

\bibitem{landweber1951iteration}
L.~Landweber, ``An iteration formula for fredholm integral equations of the first kind,'' \emph{American journal of mathematics}, vol.~73, no.~3, pp. 615--624, 1951.

\bibitem{ronneberger2015u}
O.~Ronneberger, P.~Fischer, and T.~Brox, ``U-net: Convolutional networks for biomedical image segmentation,'' in \emph{Medical Image Computing and Computer-Assisted Intervention--MICCAI 2015: 18th International Conference, Munich, Germany, October 5-9, 2015, Proceedings, Part III 18}.\hskip 1em plus 0.5em minus 0.4em\relax Springer, 2015, pp. 234--241.

\bibitem{haber2013model}
E.~Haber and M.~Holtzman~Gazit, ``Model fusion and joint inversion,'' \emph{Surveys in Geophysics}, vol.~34, pp. 675--695, 2013.

\bibitem{song2019generative}
Y.~Song and S.~Ermon, ``Generative modeling by estimating gradients of the data distribution,'' \emph{Advances in Neural Information Processing Systems}, vol.~32, 2019.

\bibitem{knoll2016joint}
F.~Knoll, M.~Holler, T.~Koesters, R.~Otazo, K.~Bredies, and D.~K. Sodickson, ``Joint mr-pet reconstruction using a multi-channel image regularizer,'' \emph{IEEE transactions on medical imaging}, vol.~36, no.~1, pp. 1--16, 2016.

\bibitem{mehranian2017synergistic}
A.~Mehranian, M.~A. Belzunce, C.~Prieto, A.~Hammers, and A.~J. Reader, ``Synergistic pet and sense mr image reconstruction using joint sparsity regularization,'' \emph{IEEE transactions on medical imaging}, vol.~37, no.~1, pp. 20--34, 2017.

\bibitem{milanfar2012tour}
P.~Milanfar, ``A tour of modern image filtering: New insights and methods, both practical and theoretical,'' \emph{IEEE signal processing magazine}, vol.~30, no.~1, pp. 106--128, 2012.

\bibitem{cho2010content}
T.~S. Cho, N.~Joshi, C.~L. Zitnick, S.~B. Kang, R.~Szeliski, and W.~T. Freeman, ``A content-aware image prior,'' in \emph{2010 IEEE Computer Society Conference on Computer Vision and Pattern Recognition}.\hskip 1em plus 0.5em minus 0.4em\relax IEEE, 2010, pp. 169--176.

\bibitem{zhang2018pet}
Y.~Zhang and X.~Zhang, ``Pet-mri joint reconstruction with common edge weighted total variation regularization,'' \emph{Inverse Problems}, vol.~34, no.~6, p. 065006, 2018.

\bibitem{choi2018pet}
J.~K. Choi, C.~Bao, and X.~Zhang, ``Pet-mri joint reconstruction by joint sparsity based tight frame regularization,'' \emph{SIAM Journal on Imaging Sciences}, vol.~11, no.~2, pp. 1179--1204, 2018.

\bibitem{leynes2018synthetic}
A.~P. Leynes and P.~E. Larson, ``Synthetic ct generation using mri with deep learning: How does the selection of input images affect the resulting synthetic ct?'' in \emph{2018 IEEE International Conference on Acoustics, Speech and Signal Processing (ICASSP)}.\hskip 1em plus 0.5em minus 0.4em\relax IEEE, 2018, pp. 6692--6696.

\bibitem{dong2020deep}
X.~Dong, Y.~Lei, T.~Wang, K.~Higgins, T.~Liu, W.~J. Curran, H.~Mao, J.~A. Nye, and X.~Yang, ``Deep learning-based attenuation correction in the absence of structural information for whole-body positron emission tomography imaging,'' \emph{Physics in Medicine \& Biology}, vol.~65, no.~5, p. 055011, 2020.

\bibitem{yang2020ct}
J.~Yang, J.~H. Sohn, S.~C. Behr, G.~T. Gullberg, and Y.~Seo, ``Ct-less direct correction of attenuation and scatter in the image space using deep learning for whole-body fdg pet: potential benefits and pitfalls,'' \emph{Radiology: Artificial Intelligence}, vol.~3, no.~2, p. e200137, 2020.

\bibitem{gong2018pet}
K.~Gong, C.~Catana, J.~Qi, and Q.~Li, ``Pet image reconstruction using deep image prior,'' \emph{IEEE Transactions on Medical Imaging}, vol.~38, no.~7, pp. 1655--1665, 2018.

\bibitem{gong2021direct}
K.~{}Gong, C.~Catana, J.~Qi, and Q.~Li, ``Direct reconstruction of linear parametric images from dynamic pet using nonlocal deep image prior,'' \emph{IEEE Transactions on Medical Imaging}, vol.~41, no.~3, pp. 680--689, 2021.

\bibitem{leynes2021attenuation}
A.~P. Leynes, S.~Ahn, K.~A. Wangerin, S.~S. Kaushik, F.~Wiesinger, T.~A. Hope, and P.~E. Larson, ``Attenuation coefficient estimation for pet/mri with bayesian deep learning pseudo-ct and maximum-likelihood estimation of activity and attenuation,'' \emph{IEEE Transactions on Radiation and Plasma Medical Sciences}, vol.~6, no.~6, pp. 678--689, 2021.

\bibitem{ho2020denoising}
J.~Ho, A.~Jain, and P.~Abbeel, ``Denoising diffusion probabilistic models,'' \emph{Advances in Neural Information Processing Systems}, vol.~33, pp. 6840--6851, 2020.

\bibitem{song2020score}
Y.~Song, J.~Sohl-Dickstein, D.~P. Kingma, A.~Kumar, S.~Ermon, and B.~Poole, ``Score-based generative modeling through stochastic differential equations,'' \emph{arXiv preprint arXiv:2011.13456}, 2020.

\bibitem{burger2014total}
M.~Burger, J.~M{\"u}ller, E.~Papoutsellis, and C.-B. Sch{\"o}nlieb, ``Total variation regularization in measurement and image space for pet reconstruction,'' \emph{Inverse Problems}, vol.~30, no.~10, p. 105003, 2014.

\bibitem{ollinger1997positron}
J.~M. Ollinger and J.~A. Fessler, ``Positron-emission tomography,'' \emph{Ieee signal processing magazine}, vol.~14, no.~1, pp. 43--55, 1997.

\bibitem{gudbjartsson1995rician}
H.~Gudbjartsson and S.~Patz, ``The rician distribution of noisy mri data,'' \emph{Magnetic resonance in medicine}, vol.~34, no.~6, pp. 910--914, 1995.

\bibitem{jack2008alzheimer}
C.~R. Jack~Jr, M.~A. Bernstein, N.~C. Fox, P.~Thompson, G.~Alexander, D.~Harvey, B.~Borowski, P.~J. Britson, J.~L.~Whitwell, C.~Ward \emph{et~al.}, ``The alzheimer's disease neuroimaging initiative (adni): Mri methods,'' \emph{Journal of Magnetic Resonance Imaging: An Official Journal of the International Society for Magnetic Resonance in Medicine}, vol.~27, no.~4, pp. 685--691, 2008.

\bibitem{schell2019automated}
M.~Schell, I.~Tursunova, I.~Fabian, D.~Bonekamp, U.~Neuberger, W.~Wick, M.~Bendszus, K.~Maier-Hein, P.~Kickingereder \emph{et~al.}, ``Automated brain extraction of multi-sequence mri using artificial neural networks,'' in \emph{Automated brain extraction of multi-sequence MRI using artificial neural networks}.\hskip 1em plus 0.5em minus 0.4em\relax European Congress of Radiology-ECR 2019, 2019.

\bibitem{yang2020model}
Y.~Yang, N.~Wang, H.~Yang, J.~Sun, and Z.~Xu, ``Model-driven deep attention network for ultra-fast compressive sensing mri guided by cross-contrast mr image,'' in \emph{Medical Image Computing and Computer Assisted Intervention--MICCAI 2020: 23rd International Conference, Lima, Peru, October 4--8, 2020, Proceedings, Part II 23}.\hskip 1em plus 0.5em minus 0.4em\relax Springer, 2020, pp. 188--198.

\bibitem{wang2004image}
Z.~Wang, A.~C. Bovik, H.~R. Sheikh, and E.~P. Simoncelli, ``Image quality assessment: from error visibility to structural similarity,'' \emph{IEEE transactions on image processing}, vol.~13, no.~4, pp. 600--612, 2004.

\bibitem{lin2017refinenet}
G.~Lin, A.~Milan, C.~Shen, and I.~Reid, ``Refinenet: Multi-path refinement networks for high-resolution semantic segmentation,'' in \emph{Proceedings of the IEEE conference on computer vision and pattern recognition}, 2017, pp. 1925--1934.

\end{thebibliography}

\vfill

\end{document}